\documentclass[twocolumn,aps,amssymb,floatfix,prl]{revtex4}
\usepackage{graphicx}
\begin{document}

\title{ Shot-noise governed Coulomb blockade in a single
Josephson junction }
\author{J. Delahaye$^1$, R. Lindell$^1$, M.S. Sillanp\"a\"a$^1$, M.A.
Paalanen$^1$, E.B. Sonin$^{1,2}$, and P.J. Hakonen$^1$}
\address{ $^1$Low Temperature Laboratory, Helsinki University of
Technology,
FIN-02015 HUT, Finland \\
$^2$The Racah Institute of Physics, The Hebrew University of
Jerusalem,
Jerusalem 91904, Israel}

\date{\today} 

\begin{abstract}
We have investigated the influence of shot noise on the IV-curves
of a single mesoscopic Josephson junction. We find that the
blockade of the Cooper pair current is strongly suppressed in the
presence of shot noise due to tunnelling in an adjacent SIN
junction. Our experimental findings can be accounted for by an
extension of the phase correlation theory. Shot-noise effect in a
resistive environment $R$ can be characterized by the effective
noise temperature $T_N=eIR /2 k_B$, which means that a Josephson
junction can easily detect shot noise from a current well below 1
pA.

\end{abstract}
 \bigskip

\maketitle

 The role of quasiparticle tunnelling on the decoherence
phenomena in Josephson junctions has become of great importance
now when schemes for quantum computation  are under development
\cite{MSS}. In the absence of dissipation from environmental
modes, shot noise provides the ultimate dephasing mechanism at low
temperatures. We have investigated the influence of external shot
noise on a single Josephson junction in a strongly resistive
environment, in which Coulomb blockade (CB) of Cooper pair current
\cite{Haviland,Penttila} takes place owing to the delocalization
of the phase variable \cite{SZ}.

The Coulomb blockade of Cooper pairs is very sensitive to
fluctuations. Inherently, it is influenced by Johnson-Nyquist
noise, which is predicted to result in a power-law-like increase
of conductance both as a function of temperature and voltage
\cite{SZ}. The exponent of the power law, $2\rho-2$, is governed
by the parameter $\rho=R/R_Q$ where $R$ describes the dissipative
ohmic environment and $R_Q=h/4e^2$.
Hence, in the case of large exponents $2\rho-2 \gg 1$, there is a
high resolution against tiny changes in temperature, or
alternatively, a high sensitivity to any external noise sources.

In this Letter we report first investigations of ``noise
spectroscopy'' using the Coulomb blockade of Cooper pairs as a
sensitive detector. We have induced shot noise by a separately
biased superconductor-insulator-normal metal (SIN) tunnel
junction. The quasiparticle current is found to strongly reduce
the CB of Cooper pairs: the influence can be resolved down to
currents of a few pA in our experiments. Our findings can be well
accounted for by including the effect of shot noise into the
phase-phase correlation functions. To our knowledge, the present
work is the first one to study quantitatively the effect of
nonlinear dissipative elements on the Cooper pair tunnelling, as
well as the first attempt to extend the phase-fluctuation theory
\cite{IN} to account for an independent shot-noise source.

When the supercurrent channel is blocked off, the current is
carried by incoherent tunnelling of Cooper pairs.
Using $P(E)$-theory \cite{IGE}, the zero-bias conductance of the
junction can be expressed via the real and the imaginary part of
$J(t)=\langle[ \varphi(t) -\varphi(0)]\varphi(0)\rangle=J_R(t)+
iJ_I(t)$:
\begin{equation}
G_0=\left.{dI\over dV}\right|_{V\rightarrow 0}=-{2 e^2E_J^2 \over
\hbar^3}\int_{-\infty}^\infty t\,dt e^{J_R(t)} \sin{J_I(t)}\,,
\end{equation}
where the real part, $J_R(t)=J_T(t)+J_N(t)$, contains the contributions
$J_T$ from the equilibrium Johnson-Nyquist noise and $J_N$ from the shot
noise. Without the shot noise
\begin{equation}
J(t)=J_T+iJ_I = 2\int_{-\infty}^\infty {d\omega \over \omega}
\frac{\mbox{Re} Z(\omega)}{R_Q}
 \frac{e^{-i\omega t} -1}{1-e^{-\beta \hbar \omega}}\,.
      \end{equation}
For ohmic environment,  $Z(\omega) =(1/R +i\omega C_T)^{-1}$ where
$C_T$ is the junction capacitance. The imaginary part $J_I(t)=-\pi
\rho\left(1-e^{-|t|/\tau}\right)\mbox{sign}t$ does not depend on
temperature and in the low-temperature limit $\hbar \beta
=\hbar/k_BT \gg \tau$,
\begin{eqnarray}
J_T(t) =  2\rho \bigg\{{1\over
2}\left[e^{-t/\tau}\mbox{Ei}\left({t\over \tau}\right)+
e^{t/\tau}\mbox{Ei}\left(-{t\over \tau}\right) \right] \nonumber \\
-\ln\left[{\hbar \beta \over\pi \tau}\sinh\left({\pi t \over \hbar
\beta}\right)\right]
 -\gamma \bigg\}\,,
      \end{eqnarray}
 where $\tau=RC_T$. The temperature dependence is important only at long
times $t \gg \hbar \beta \gg \tau $, where temperature determines
the phase diffusion, and $J_R(t) \approx - 2\pi\rho  t /\hbar
\beta=- 2\pi\rho k_BT t /\hbar$. Now we consider the contribution
$J_N$ of the shot noise. From the current--current spectral
density $S_I =2eI$ one can find the voltage-voltage spectral
density:
\begin{equation}
S_V = |Z(\omega)|^2 S_I= {2eI R^2 \over 1+(\tau \omega)^2}~.
      \end{equation}
Then using the Josephson relation $\hbar {\partial \varphi /
\partial t}= 2e V$ one can obtain the phase-phase spectral density
$S_\varphi=(4e^2 /\hbar^2 \omega^2)S_V$ and finally the
contribution of the shot noise to $J_R$ is
\begin{eqnarray}
J_N(t) ={\pi I \over e}{R^2 \over R_Q^2}\int_{0}^\infty
{d\omega\over \omega^2}{\cos \omega t -1 \over 1+(\tau \omega)^2}
\nonumber \\
 = -{\pi^2 I \tau \over 2 e}{R^2 \over R_Q^2}
\left(e^{-|t|/\tau} +{|t|\over \tau} -1\right)~.
     \label{J-s} \end{eqnarray}
As well as the thermal contribution, for small $I$ the shot-noise
contribution is significant only for large $t \gg \tau$ where the
shot noise modifies the phase diffusion and $J_T(t)+J_N(t)
=-2\pi\rho k_B( T +T_N)t /\hbar$, where $T_N =eIR/2k_B$ is the
noise temperature.

Using the asymptotic expressions for $J_T(t)$ and $J_N(t)$ at $t
\gg \tau$, one can derive an analytical expression for the
conductance:
\begin{eqnarray}
G_0={2 e^2 E_J^2 \over \hbar^3 }\sin(\pi \rho)e^A e^{-2\rho
\gamma_E} \left(2\pi \tau \over\hbar\beta \right)^{2\rho-2} \tau^2
\nonumber \\
\times{d\mbox{B}(a/2+\rho,1-2\rho)\over da} \, ,
\label{G}\end{eqnarray} where $A=2\pi k_BT_N\rho /\hbar$, $a=2\rho
T_N/T$, $\gamma_E=0.577...$ is the Euler constant and
$\mbox{B}(x,y)=\Gamma(x)\Gamma(y)/\Gamma(x+y)$ is the beta
function. Using the functional relations between gamma functions,
one can check that in the limit $T_N \rightarrow 0$ $(a
\rightarrow 0)$, Eq. (\ref{G}) coincides with the conductance
\begin{eqnarray}
G_0={2\pi e^2 E_J^2 \over \hbar^3 }e^{-2\rho \gamma} \left(2\pi
\tau \over\hbar\beta \right)^{2\rho-2} \tau^2{\Gamma(\rho)\over
\Gamma(2\rho)} \propto T^{2\rho -2} \label{GT}\end{eqnarray}
calculated in Ref. \cite{IGE} without the shot noise. In the
opposite limit $T_N \gg T$ ($a \rightarrow \infty$) one can use
the asymptotic relation $\Gamma(z+y) \sim \Gamma(z)z^y$ at
$z\rightarrow \infty$, and Eq. (\ref{G}) yields
\begin{eqnarray}
G_0={4\pi \rho \tau^2 e^2 E_J^2 \over \hbar^3 } \left(2\pi \rho
\tau k_B\over\hbar \right)^{2\rho-2} (T+T_N)^{2\rho -2}  \,
   \label{GN}        \end{eqnarray}
for small $\rho$. In contrast to Eq. (\ref{GT}), Eq. (\ref{G})
cannot be extended to large $\rho$, where the Coulomb blockade
takes place and our experiment has been done. This is because
times $t \sim \tau$ become relevant and the oscillating term
$\sin(\pi \rho)$ is not cancelled at $a \neq 0$. But one may
expect that the effect of shot noise still can be accounted for by
the expression $G \sim (T+\gamma(\rho) T_N)^{2\rho -2}$, where
$\gamma(\rho)$ may be considered as a $\rho-$dependent Fano-factor
of order 1.

\begin{figure}

    \includegraphics[width=7cm]{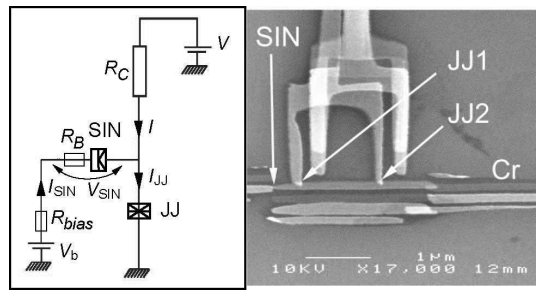}

    \caption{A scanning electron microscope picture of sample 1 and
    a schematic view of the circuit. The chrome resistor is denoted by
    Cr, the superconductor-normal junction by SIN, and the Josephson
    junction in a SQUID-loop configuration by JJ1 and JJ2. For
    constructional details, see text. }\label{SEM}

    \end{figure}

Our experiments were performed using a circuit layout which is
 depicted in Fig. \ref{SEM}.  The physical
structure consists of four basic elements: 1) an Al-AlO$_x$-Al
Josephson junction (JJ) with a tunnel resistance of $R_T^{JJ} =
4-8$ k$\Omega$, 2) a superconducting-normal Al-AlO$_x$-Cu tunnel
junction (SIN) with $R_T^{SIN} = 6-27 $ k$\Omega$, 3) a thin film
Cr resistor of $R_C = 23-67 $ k$\Omega$ (20 $\mu$m long), located
within a few $\mu$m from the Josephson junction, and 4) a similar
Cr resistor $R_B$ in front of the SIN junction. Altogether we
investigated three samples, the parameters of which are given in
Table I.

The circuits were fabricated using electron beam
lithography and four-angle evaporation. The Cr resistor (5 nm
thick, 100 nm wide) was evaporated first at an angle of
$-18^{\circ}$, followed by the Al-island at $-38^{\circ}$. After
oxidation, the sample holder was rotated by $45^{\circ}$ around
the z-axis and the SQUID-loop was deposited by a second
Al-evaporation at $+38^{\circ}$. Finally, the SIN-junction was
made in copper deposition at $+6^{\circ}$.

The JJ junction was, in fact, made of two $100 \times 100$ nm$^2$
junctions in a SQUID geometry in order to facilitate tuning of its
Josephson energy. The Josephson energy $E_J^{max}$ at no magnetic
flux was calculated from the tunnelling resistance using the
Ambegaokar-Baratoff relation. The minimum Josephson coupling
energy, $E_J^{min}$, was obtained from the minimum of critical
current $I_C(\Phi)$ as a function of external flux $\Phi$, and
assuming a linear dependence between $E_J$ and $I_C$. The Coulomb
energy $E_C=e^2/2C$ was estimated from the IV-curves in the normal
state: the sum of junction capacitances, $C=C_{SIN}+C_{JJ}$, was
obtained from the voltage offset at large bias voltages using the
formula $V_{\mathrm{offset}}=\frac{e}{2C}$. The ratio $E_J$/$E_C$
could be tuned over the range 0.33 and 4.3 (see Table I). External
noise was filtered out by 1.5 MHz low-pass filters at the top of
the cryostat and by 1 m of Thermocoax cable at the mixing chamber.

\begin{figure}

    \includegraphics[width=7cm]{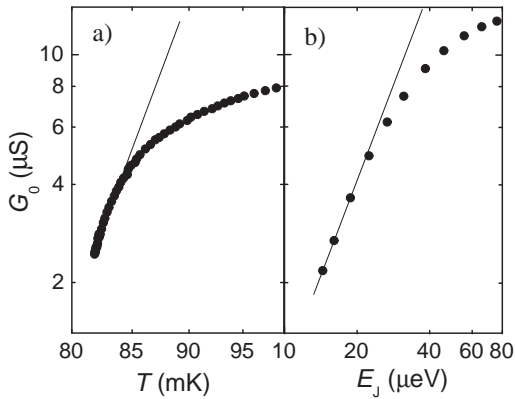}

    \caption{a) Zero bias conductance $G_0=\frac{dI_{JJ}}{dV}|_{I_{JJ}=0}$ for the
    JJ + $R_C$ section of sample 3 as a function of temperature $T$ at $E_J=14$ $\mu$eV. The solid
    curve illustrates Eq. (\ref{GT}) using the ohmic environment of
    $R=R_C$. b) $G_0$ as a function of Josephson energy
    $E_J$ at $T=82$ mK. The solid curve illustrates the quadratic dependence
    obtained from Eq. (\ref{GT}). } \label{JJ T}
    \end{figure}

\begin{table}
\begin{tabular}{|c|c|c|c|c|c|c|}
   \hline  & $R_T^{JJ}($k$ \Omega )$ & $R_T^{SIN}($k$ \Omega )$& $R_{C}($k$ \Omega )$ &
                  $R_{B}($k$ \Omega )$ & $E_C$  & $E_J^{min}$ / $E_J^{max}$ \\
   \hline 1 & 8.1 & 27.3  & 22.6 & 0.1 & 65 & 22 / 78 \\
   \hline 2 & 7.8 & 5.8  & 54.2 & 0.1 & 50 & 83 / 83 \\
   \hline 3 & 4.3 & 10 & 67 & 53 & 35 & 14 /150  \\

      \hline
  \end{tabular}
  \caption{Device parameters for our three samples numbered
  consecutively by the first column.
  The next two columns give the
  tunnelling resistance of the Josephson junction $R_T^{JJ}$ and
  the SIN junction $R_T^{SIN}$. $R_{B}$ and $R_{C}$ denote the
  impedances in the immediate vicinity of the SIN and Josephson junctions, respectively.
  The last two columns indicate the Coulomb energy, $E_C$, and the
  minimum $E_J^{min}$ and maximum $E_J^{max}$ values of the Josephson
  energy. The energies are given in $\mu$eV.}

\end{table}
Fig. \ref{JJ T}a displays the temperature dependence of zero-bias
conductance $G_0(T) \propto T^{(2\rho-2)}$, measured on sample 3.
Using $R=R_C=67$ k$\Omega$, we get for the exponent $2\rho-2=19$
which yields the power law shown by the solid curve in Fig.
\ref{JJ T}. This verifies that the steep, measured $G_0(T)$ of our
sample agrees quite well with the theoretical temperature scaling
law of Eq.~(\ref{GT}) at the lowest temperatures. Slightly better
agreement was found for sample 1 where the condition for the
validity of the theory, $\rho k_B T<E_C$, is easier to fulfill.

As a second test of theory, we show the dependence of $G_0$ on the
Josephson energy $E_J$ in Fig. \ref{JJ T}b. At low values of $E_J$
we recover the expected $E_J^2$ dependence. Hence, we conclude
that, in the small $E_J$ limit, the conduction in the JJ is
predominantly caused by inelastic Cooper pair tunnelling, and no
extra leakage is present at the lowest temperatures. Under these
circumstances, any change in $G_0$ can be assigned to an
additional external source of noise in the circuit.

\begin{figure}

    \includegraphics[width=7cm]{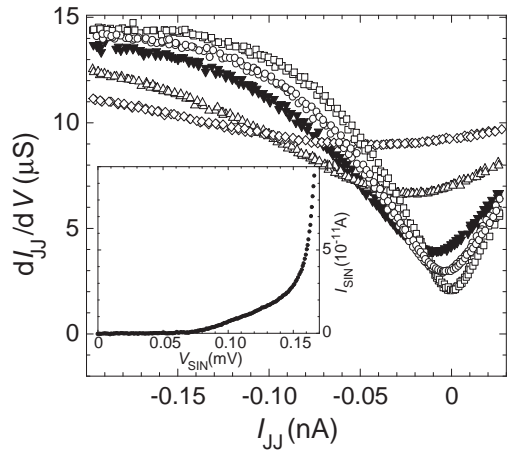}

    \caption{Differential conductance $\frac{dI_{JJ}}{dV}$ vs. current
    $I_{JJ}$ for the JJ + $R_C$ section of sample 3 with a small current bias (via
    100
    M$\Omega$ resistor) in the SIN junction: $I_{SIN}=$ 0 ($\square$), 0.01
    ($\circ$), 0.02 ($\blacktriangledown$), 0.05 ($\vartriangle$), 0.1
    nA ($\lozenge$). The inset shows the $I_{SIN}$ vs. $V_{SIN}$ for
    the SIN-junction biased via JJ ($R_C$ was employed as the voltage
    lead). $T=82$ mK.} \label{JJ RI}
    \end{figure}

The Coulomb blockade at small voltages is seen most clearly using
measurements of differential conductance $\frac{dI}{dV}$. Fig.
\ref{JJ RI} displays the measured $\frac{dI_{JJ}}{dV}$ for sample
3 at zero quasiparticle current as well as at a few values of
$I_{SIN}$ ranging from 0.01 nA to 0.1 nA; the data was taken at
the minimum value of $E_J=14$ $\mu$eV. The bias voltage $V$ to the
JJ was applied via the chrome resistor while the SIN-junction was
current biased through $R_{bias}=100$ M$\Omega$. The minimum
amplitude of the Coulomb blockade dip $G_{min}=\left[
\frac{dI_{JJ}}{dV}\right]_{min}$ is seen to increase monotonically
with $I_{SIN}$. In addition, there is a small shift by $\Delta
I_{JJ} \sim -0.20 \cdot I_{SIN}$ in the location of the minimum
conductance with increasing $I_{SIN}$. Sample 1 yielded $\Delta
I_{JJ} \sim -0.22 \cdot I_{SIN}$, but in sample 2, the large value
of $E_J$ prevented any quantitative analysis.

The IV-curve ($I_{SIN}$ vs. $V_{SIN}$, see Fig. \ref{SEM}) of the
SIN junction is illustrated in the inset of Fig. \ref{JJ RI}. The
biasing was applied through JJ while the voltage was recorded via
$R_C$. In the subgap region, there is only a small current,
presumably due to Andreev reflection processes \cite{Andreev}. At
the gap edge, the current increases rapidly and the differential
resistance $R_{d}=\frac{dV_{SIN}}{dI_{SIN}}$ drops down to 5 - 10
k$\Omega $ at currents 0.05 nA - 1 nA. This means that the
resistive environment $R$ seen by the Josephson junction varies
with the biasing of the SIN junction. When $I_{SIN}< 20-30 $ pA,
$R_{SIN}\gg R_C$ and the environment is purely governed by $R_C$.
On the other hand, in the regime  $0.05 - 1$ nA, we may
approximate  $R^{-1} \sim R_C^{-1} + (R_B+R_d)^{-1}$. Here, we
neglect all the second order terms which might give a noticeable
contribution to the dissipative part of the impedance
\cite{Gupta}.

In addition to shot noise, the SIN junction might give a
contribution via the Johnson--Nyquist noise, which is determined
by the differential resistance $R_d$ at the biasing point. Since
$R_d$ is not monotonous and has a minimum as a function of
$I_{SIN}$, this would result in a nonmonotonous current dependence
for the Coulomb blockade. None of our samples, however, showed any
nonmonotonous behavior. The absence of any re-entrant type of
behavior supports our observation that we are dealing with the
effect of shot noise.

\begin{figure}

        \includegraphics[width=7cm]{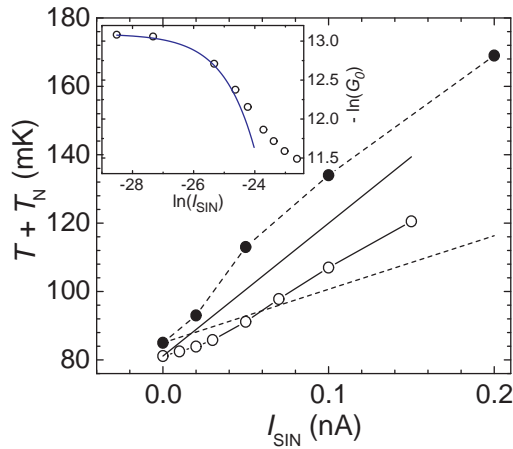}

    \caption{Temperature $T+T_{N}$ vs. $I_{SIN}$
    obtained by comparing $G_0(I_{SIN})$ with $G_0(T)$ : sample 1
    ($\bullet ---$) and sample 3
    ($\circ$ -----). The dashed and solid lines illustrate the formula
    $eI_{SIN}R/2k_B$ for samples 1 and 3, respectively. Inset: $G_0$
    of sample 3 as a function of $I_{SIN}$. The fit is obtained using
    Eq. (\ref{GN}) with $R^{-1}=R_C^{-1} + R_B^{-1}$ in order to have
    a better environment for the high current regime.} \label{JJ ISN}
    \end{figure}

As there are uncertainties in the theory, we deduce $T_N$ by
equating $G_0(I_{SIN})$ with $G_0(T+T_N)$. The results are
displayed in Fig. \ref{JJ ISN} for samples 1 and 3. Both samples
show a nearly linear increase in $T_N$ with growing $I_{SIN}$. A
comparison with the formula $T_N =eIR/2k_B$, yields for the
Fano-factor $\gamma=3$ and 0.5 for the samples 1 and 3,
respectively. The inset of Fig. \ref{JJ ISN} displays the
$I_{SIN}$-dependence of $G_0$ for sample 3. A fit using Eq.
(\ref{GN}) with a "high-current" environment of $R^{-1}=R_C^{-1} +
R_B^{-1}$ yields $\gamma \sim 1$.

The measured Fano-factors, at least qualitatively, agree with our
theoretical scenario, which up to now has not yet provided
numerical values for the Fano factor at large $\rho$. But there
are additional, neglected physical processes, which also can
account for the deviation of the measured Fano-factors from 1
\cite{Blanter}. Especially, there is an uncertainty in the subgap
regime of the SIN junction: the conduction should be caused only
by Andreev processes. This would indicate that the Fano factor
should be 2 for $I_{SIN}$ (tunnelling of 2e charges). At larger
currents, the dynamic impedance of the SIN junction is expected to
lower the exponent $\rho$. Last but not least, there is the
question of correlations between quasiparticle and Cooper pair
tunnelling.

Our model assumes additivity of thermal and shot noise and,
therefore, cannot account for the shift of the minimum of
conductance to finite current values by $\Delta I_{JJ}$ in Fig.
\ref{JJ RI}. But one should not rule out the possibility of two
correlated noise sources, one on SIN and one on JJ.
Phenomenologically, we may write

\begin{equation}\label{pheno}
  T_{eff}=T+\frac{eR}{2k_B}\left[\gamma I_{SIN}+\kappa 2I_{JJ}+\lambda
  \sqrt{2\gamma\kappa I_{SIN}I_{JJ}}\right]
\end{equation}
where $\kappa$ denotes the Fano factor of the Cooper pair
tunnelling noise in the Josephson junction and $\lambda$ describes
the correlations between the two noise sources. By minimizing this
with respect to $I_{JJ}$ at a fixed $I_{SIN}$, we obtain
\begin{equation}\label{equil}
  I_{JJ}=\frac{\gamma \lambda^2}{8\kappa}I_{SIN}
\end{equation}
which defines the condition for the minimum $T_{eff}$ (with
$\lambda<0$).

Using the experimental value of $\gamma$ and the shift $\Delta
I_{JJ}$ for sample 3, we get an upper limit $\kappa<0.31$ (at
maximum correlation $\lambda=-1$). We expect this limit to hold
for sample 1 as well: then, by taking $\kappa=0.31 $, we obtain
$\lambda=-0.43$. Thus, it appears that the correlations between
$I_{JJ}$ and $I_{SIN}$ depend on the base resistor and that an
increase in $R_B$ tends to suppress uncorrelated tunnelling
current in the SIN junction.

According to our experimental analysis, a single Josephson
junction provides a good candidate for a noise detector in
general. Its main virtue is the high sensitivity which comes from
the large detector band width: $\sim 1/RC$. Experimentally,
changes by 1 mK in $T_{N}$ can be resolved clearly in Fig. \ref{JJ
ISN}. This corresponds to a quasiparticle current of $I_{qp}=3$
pA, which equals the sensitivity in high-resolution noise
experiments of Ref. \cite{Reznikov}. An enhancement by a factor of
ten in the noise sensitivity is obtainable by improving the
stability of $\frac{\Delta R}{R}$ measurement to the level of
$1\%$ and working at $T=20$ mK. This would allow, for example,
detailed studies of the back-action noise of quantum amplifiers
such as superconducting SETs. A larger dynamic range can be
obtained by operating the device at lower $T$.

In summary, our measurements of $G$ vs. $I$ curves for solitary,
resistively confined small Josephson junctions show that the
Cooper pair blockade can be suppressed strongly by shot noise from
a near-by SIN tunnel junction. Using the framework of the phase
correlation theory, we have presented a theoretical analysis of
the shot-noise effect, which can be characterized by an effective
noise temperature of the system $T_N$. This approach yields a good
agreement with our experimental results. We expect that other
sources of noise ($1/f$--noise, as an example), produce similar
effects. Consequently, CB of Cooper pairs can be employed as a
sensitive noise detector with a resolution of 0.1 mK in $T_N$.

We acknowledge fruitful discussions with L. Roschier, A. Schakel
and A. Zaikin. This work was supported by the Academy of Finland
and by the Large Scale Installation Program ULTI-3 of the European
Union. A research grant of the Israel Academy of Sciences and
Humanities is gratefully acknowledged by EBS.

\end{document}